\documentclass[aip,jcp,amsmath,amssymb,reprint,floatfix]{revtex4-2}

\usepackage{graphicx}
\usepackage{bm}
\usepackage{mathtools}
\usepackage{microtype}
\usepackage[T1]{fontenc}
\usepackage[utf8]{inputenc}
\usepackage{mathptmx}
\usepackage[hidelinks]{hyperref}

\begin{document}

\title{Chiral Torsional Double Wells: Tunneling and Polarizability}

\author{Shah Ishmam Mohtashim}
\affiliation{Department of Chemistry, North Carolina State University,
Raleigh, North Carolina 27695, United States}

\author{Manas Sajjan}
\affiliation{National Center for Computational Sciences, Oak Ridge National
Laboratory, Oak Ridge, Tennessee 37830, United States}

\author{Sabre Kais}
\email{skais@ncsu.edu}
\affiliation{Department of Chemistry, North Carolina State University,
Raleigh, North Carolina 27695, United States}
\affiliation{Department of Electrical and Computer Engineering, North
Carolina State University, Raleigh, North Carolina 27695, United States}

\date{\today}

\begin{abstract}
Quantum tunneling between mirror-related wells of a torsional potential sets the timescale for coherent interconversion of chiral conformers. We ask whether slower tunneling also weakens the transition polarizability in hydrogen peroxide, hydrogen disulfide, and hydrogen diselenide. We use relaxed CCSD(T)//MP2 potentials and a kinetic model that removes overall molecular rotation to obtain the torsional states, then use static polarizabilities to calculate transition matrix elements between them. In the molecular frame used here, symmetry makes the matrix element of the even dipole component vanish between opposite-parity torsional states, while the odd polarizability component has a nonzero matrix element. The two quantities probe different parts of the torsional landscape: tunneling depends on penetration through the barriers, while the polarizability matrix element forms mainly within the wells. Consequently, tunneling slows sharply across the molecular series even as the matrix element grows. Higher torsional doublets and deuterium substitution further show that large changes in tunneling time can occur with little change in the transition polarizability matrix element. Calculated torsional energies for H$_2$O$_2$ and D$_2$O$_2$ agree with experimental values. Thus, slowing coherent interconversion need not weaken the intrinsic transition polarizability.

\end{abstract}

\maketitle

\section{Introduction}\label{sec:intro}

H$_2$O$_2$ and its heavier analogues H$_2$S$_2$ and H$_2$Se$_2$ are among the simplest molecules whose enantiomeric conformers interconvert by
internal rotation. Each has two mirror-related, nonplanar minima along the H--X--X--H torsion ($X=\mathrm{O,S,Se}$). For a symmetric torsional
Hamiltonian, the even and odd eigenstates extend over both wells. Appropriate real, equal-amplitude superpositions of the lowest pair are localized mainly in opposite wells. Their energy splitting determines the coherent transfer time between these localized states. \cite{Hund1927,Quack2002,Prentner2015}

The two wells are separated by planar cis and trans geometries. Spectroscopy establishes a ground-doublet splitting of hundreds of gigahertz in H$_2$O$_2$, while calculations predict splittings in the
kilohertz range for the heavier analogues. \cite{Redington1962,Flaud1989,Sahu2021} These frequency scales correspond to coherent transfer times spanning
nearly seven orders of magnitude. In the molecular-frame convention
defined in Sec.~\ref{subsec:frame}, $\mu_x$ is even under well exchange
and $\alpha_{yz}$ is odd.

Grishanin and Zadkov analyzed dipole cancellation in a torsional model of H$_2$O$_2$ and explored an electric-quadrupole route to light-induced chirality.\cite{Grishanin1999}
Pitsevich and co-workers calculated the H$_2$O$_2$ rotation--torsion Raman spectrum using HF/6-311G polarizabilities, explicit molecular-frame continuation, and rotational matrix elements.\cite{Pitsevich2010}
Multidimensional studies of H$_2$O$_2$ stereomutation and of isotope effects on tunneling in the sulfur and selenium analogues show that the splitting is highly sensitive to mass.\cite{Fehrensen1999,Fehrensen2007,Gottselig2001,Gottselig2003}
Previous studies establish tunneling splittings across the hydrogen
dichalcogenides and Raman spectra for H$_2$O$_2$.
Here we quantify how a torsional polarizability matrix element varies
relative to the splitting across chemical substitution, torsional
excitation, and isotope substitution.

We calculate torsional eigenstates on relaxed CCSD(T)//MP2 potentials
with a Fourier discrete-variable representation (DVR). The kinetic
metric excludes overall rotation, and static PBE0 polarizabilities
provide the electronic response along the path. We compare the
torsional levels with spectroscopy and previous calculations, then
examine the splitting and polarizability matrix element across the
O/S/Se series. Two comparisons retain the electronic surfaces: the
three lowest parity pairs of H$_2$S$_2$ and H$_2$Se$_2$, and the
H$_2$O$_2$/D$_2$O$_2$ isotope pair. These comparisons distinguish changes
in the nuclear states from changes in the electronic response.

\section{Theory and computational methods}\label{sec:methods}

\subsection{Relaxed torsional path}\label{subsec:path}

The H--X--X--H dihedral angle $\phi$ defines the torsional path: cis lies at $\phi=0$, trans at $\phi=\pm\pi$, and the reflection $\phi\to-\phi$ exchanges the two torsional wells.
At each angle we optimize, under $C_2$ symmetry, the X--X distance, the two equal X--H distances, and the two equal H--X--X angles.
Because we omit the electroweak parity-violating interaction, which would make the two wells very slightly inequivalent,\cite{Quack2002,Laerdahl1999,QuackSeyfangWichmann2022} mirror-image geometries have equal energies:
\begin{align}
V(-\phi)&=V(\phi).
\label{eq:Vparity}
\end{align}
All parity labels and tunneling times refer to this symmetric Hamiltonian.\cite{Gottselig2003}

We optimize geometries with second-order M{\o}ller--Plesset perturbation theory (MP2) and compute path energies with coupled-cluster singles, doubles, and perturbative triples [CCSD(T)]; we denote this combination CCSD(T)//MP2.\cite{Moller1934,Purvis1982,Raghavachari1989}
All electronic-structure calculations use PySCF.\cite{Sun2018,Sun2020}
Hydrogen, oxygen, and sulfur use the aug-cc-pVTZ basis; selenium uses aug-cc-pVTZ-PP with its small-core pseudopotential.\cite{Dunning1989,Kendall1992,Peterson2003}
The MP2 and CCSD(T) calculations freeze two occupied orbitals in H$_2$O$_2$ and ten in H$_2$S$_2$; in H$_2$Se$_2$, the pseudopotential replaces the core and no further orbitals are frozen. 
The chemical series uses the standard atomic weights as masses: $1.00784$ (H), $15.999$ (O), $32.06$ (S), and $78.971$~amu (Se).
The isotope comparison (Sec.~\ref{subsec:isotope}) uses isotopic masses.

We sample $\phi$ from $0^\circ$ to $180^\circ$ in $5^\circ$ steps, refined to $2.5^\circ$ near the planar geometries and the torsional minima; Eq.~\eqref{eq:Vparity} gives the negative branch.
For the chemical series, a periodic cubic spline interpolates the energies, and cubic splines with zero slope at $0$ and $\pi$ interpolate the relaxed internal coordinates.

Multidimensional treatments of H$_2$O$_2$ include the other nuclear motions.\cite{Kuhn1999,Fehrensen1999,Fehrensen2007}
Our one-dimensional relaxed path treats the three molecules uniformly but omits the torsional dependence of the zero-point energy of the other modes and their dynamical coupling to the torsion. Agreement with experiment tests the model as a whole and cannot isolate this error.

\subsection{Torsional kinetic energy}\label{subsec:metric}

We describe torsion along the relaxed one-dimensional path after removing
overall translation and rotation.\cite{MillerHandyAdams1980}
Let $\mathbf r_a(\phi)$ denote the center-of-mass-referenced position of
atom $a$, and let $\mathbf u_a(\phi)=\partial\mathbf r_a/\partial\phi$
be its displacement per unit change in dihedral angle. Part of this
displacement may correspond to a rigid rotation of the entire molecule.
The mass-weighted best-fitting rotation per unit angle satisfies
\begin{align}
\mathbf J(\phi)\bm\omega(\phi)
&=\sum_a m_a\,\mathbf r_a(\phi)\times\mathbf u_a(\phi),
\label{eq:omega}\\
\mathbf J(\phi)
&=\sum_a m_a\left[
|\mathbf r_a(\phi)|^2\mathbf 1
-\mathbf r_a(\phi)\mathbf r_a^{\mathsf T}(\phi)
\right],
\label{eq:inertia}
\end{align}

where $m_a$ is the mass of atom $a$, $\mathbf J$ is the inertia tensor, and $\mathbf 1$ is the $3\times3$ identity. Removing this rotation gives the internal displacement and the effective moment of inertia,
\begin{align}
\mathbf u_a^{\mathrm{int}}(\phi)
&=\mathbf u_a(\phi)-\bm\omega(\phi)\times\mathbf r_a(\phi),\\
I_{\mathrm{eff}}(\phi)
&=\sum_a m_a\left|\mathbf u_a^{\mathrm{int}}(\phi)\right|^2.
\label{eq:Ieff}
\end{align}

$I_{\mathrm{eff}}(\phi)$ is the mass-weighted squared internal displacement per unit change in $\phi$. It varies along the path, and the kinetic energy of the one-dimensional model is $T=\tfrac12 I_{\mathrm{eff}}(\phi)\dot\phi^2$.

For the numerical derivative, we independently align the geometries at
$\phi\pm2\times10^{-4}$~rad to the central geometry using a
mass-weighted Kabsch procedure, then take their central difference.\cite{Kabsch:a12999}
We also calculate Eq.~\eqref{eq:Ieff} by projecting rigid rotation
from the unaligned tangent. 

To use a uniform periodic grid, we absorb the angle dependence of $I_{\mathrm{eff}}$, which is even in $\phi$, into a new coordinate: the signed mass-weighted arc length, rescaled so that one torsional period spans $\chi\in[-\pi,\pi)$,
\begin{align}
s(\phi)&=\int_0^\phi\sqrt{I_{\mathrm{eff}}(u)}\,du,
\label{eq:arc_length}\\
\chi(\phi)&=\pi\frac{s(\phi)}{s(\pi)},&
I_{\mathrm{arc}}&=\left[\frac{s(\pi)}{\pi}\right]^2.
\label{eq:arc}
\end{align}
For a given molecule and choice of nuclear masses, $s(\pi)$ is
evaluated at the fixed endpoint $\pi$, so $I_{\mathrm{arc}}$ is
a constant. Its value can differ between molecules and
isotopologues. The angle-dependent inertia
$I_{\mathrm{eff}}(\phi)$ determines the nonuniform mapping
between $\phi$ and $\chi$. In particular,
\begin{equation}
\frac{d\chi}{d\phi}
=\sqrt{\frac{I_{\mathrm{eff}}(\phi)}{I_{\mathrm{arc}}}},
\qquad
\frac12 I_{\mathrm{eff}}(\phi)\dot\phi^2
=\frac12 I_{\mathrm{arc}}\dot\chi^2.
\end{equation}
The coordinate transformation therefore preserves the kinetic energy while giving it a constant coefficient.

We quantize the reduced path model in the arc coordinate.
In atomic units, the torsional Hamiltonian is
\begin{align}
H_{\mathrm{tor}}
&=-\frac{1}{2I_{\mathrm{arc}}}
  \frac{\partial^2}{\partial\chi^2}
  +V[\phi(\chi)].
\label{eq:Htor}
\end{align}
The derivative acts on the torsional wavefunction; the constant
$I_{\mathrm{arc}}$ sets the kinetic-energy scale. Here $V[\phi(\chi)]$ is the relaxed potential at the geometry that corresponds to $\chi$, so the change of coordinate leaves the potential unchanged. 
Wavefunctions are normalized with respect to $d\chi$. A probability
density reported per unit physical dihedral angle includes the
Jacobian $d\chi/d\phi$.

For each molecule, we evaluate $I_{\mathrm{eff}}$ on the spline-interpolated geometries at 4097 angles on $[0,\pi]$. Cumulative trapezoidal integration gives $s(\phi)$; shape-preserving cubic (PCHIP) interpolation gives $I_{\mathrm{eff}}(\phi)$ and the inverse map $\phi(\chi)$; and reflection extends the map to negative angles. Tables report $I_{\mathrm{arc}}$ in amu\,bohr$^2$.

\subsection{Parity states and tunneling times}\label{subsec:dvr}

We solve Eq.~\eqref{eq:Htor} with a periodic Fourier DVR\cite{Light1985,Colbert1992} on $N=384$ uniformly spaced $\chi$ points.
We diagonalize the reflection-even and reflection-odd blocks separately, which prevents numerical mixing of nearly degenerate states.

Let $|+\rangle$ and $|-\rangle$ be the lowest even and odd states, with energies $E_+$ and $E_-$.
We use real wavefunctions and choose their relative sign so that $\int_0^\pi\psi_+(\chi)\psi_-(\chi)\,d\chi>0$.
The splitting, expressed as an energy, a wavenumber, and a frequency, is
\begin{align}
\Delta E_{01}&=E_--E_+, &
\Delta_{01}&=\frac{\Delta E_{01}}{hc}, &
\nu_{01}&=c\Delta_{01},
\label{eq:splitting}
\end{align}
with $\Delta_{01}$ in cm$^{-1}$ and $c$ in cm\,s$^{-1}$.
The combinations
\begin{align}
|\mathrm L\rangle=\frac{|+\rangle-|-\rangle}{\sqrt2},\qquad
|\mathrm R\rangle=\frac{|+\rangle+|-\rangle}{\sqrt2}.
\label{eq:LR}
\end{align}
are localized mainly in opposite torsional wells.
Because the even and odd states have different energies, their
relative phase changes with time, shifting the probability from
one torsional well to the other.
With this convention, \(|\mathrm L\rangle\) is localized mainly at \(\phi<0\) and \(|\mathrm R\rangle\) at \(\phi>0\), as labeled in Fig.~\ref{fig:landscapes}. The labels do not specify absolute stereochemical configurations.
Starting from $|\mathrm L\rangle$, the probability of finding
$|\mathrm R\rangle$ at time $t$, and the time of its first maximum, are
\begin{align}
P_{\mathrm R}(t)
&=|\langle\mathrm R|e^{-iH_{\mathrm{tor}}t/\hbar}
|\mathrm L\rangle|^2=\sin^2(\pi\nu_{01}t),\\
t_{\mathrm L\rightarrow\mathrm R}&=\frac{1}{2\nu_{01}}.
\label{eq:fieldfree}
\end{align}
We call $t_{\mathrm L\rightarrow\mathrm R}$ the first coherent well-to-well transfer time, or tunneling time for short.
It describes coherent motion in an isolated molecule, not irreversible racemization.

Figure~\ref{fig:landscapes} labels the six lowest torsional
eigenstates by $v=0,\ldots,5$ in order of increasing energy.
We index the three adjacent even--odd pairs by $n=0,1,2$,
with $|n,+\rangle\equiv|v=2n\rangle$ and
$|n,-\rangle\equiv|v=2n+1\rangle$.
Thus, the $n=0$ pair is the lowest doublet defined above.
For each pair we compute the splitting
$\Delta_n=(E_{n,-}-E_{n,+})/(hc)$, the transition matrix element
$A_n^{yz}=\langle n,+|\hat\alpha_{yz}|n,-\rangle$, and the
squared coupling $S_n^{yz}=|A_n^{yz}|^2$, so that
$\Delta_0=\Delta_{01}$ and $A_0^{yz}=A_{01}^{yz}$.
The ratio $S_n^{yz}/S_0^{yz}$ compares pairs within one molecule.
When pair $n$ supports well-localized superpositions, $t_n=1/(2c\Delta_n)$ is its tunneling time, with $t_0=t_{\mathrm L\rightarrow\mathrm R}$.

To test localization, we form the equal superposition of $|n,+\rangle$ and $|n,-\rangle$ with the largest probability at $\phi>0$ and call that probability $P_{\mathrm{well},n}$.
By symmetry, the opposite-sign superposition has the same probability at $\phi<0$.
Values near one mean that the pair supports well-localized states.

\subsection{Body-fixed frame and electronic properties}\label{subsec:frame}

We express the electric dipole and polarizability in a molecular
frame that follows the relaxed torsional path. The electronic-structure
output uses fixed Cartesian axes
$\{\hat{\mathbf e}_X,\hat{\mathbf e}_Y,\hat{\mathbf e}_Z\}$.
At each angle $\phi$, we direct $\hat{\mathbf z}$ from X$_1$ to
X$_2$, choose $\hat{\mathbf x}$ along the $C_2$ symmetry axis
perpendicular to the X--X bond, and set
$\hat{\mathbf y}=\hat{\mathbf z}\times\hat{\mathbf x}$.

We place these molecular unit vectors, written in fixed Cartesian
coordinates, in the columns of $B(\phi)$:
\begin{equation}
B(\phi)=
\bigl(\hat{\mathbf x}(\phi)\ 
      \hat{\mathbf y}(\phi)\ 
      \hat{\mathbf z}(\phi)\bigr),
\qquad B(\phi)^{\mathsf T}B(\phi)=\mathbf I_3.
\end{equation}
The transpose $B(\phi)^{\mathsf T}$ projects the Cartesian
electronic-structure output onto the molecular axes:
\begin{equation}
\begin{aligned}
\bm\mu^{\mathrm{BF}}(\phi)
&=B(\phi)^{\mathsf T}\bm\mu^{\mathrm{cart}}(\phi),\\
\bm\alpha^{\mathrm{BF}}(\phi)
&=B(\phi)^{\mathsf T}\bm\alpha^{\mathrm{cart}}(\phi)B(\phi).
\end{aligned}
\label{eq:BFtransform}
\end{equation}
For example,
$\alpha_{yz}^{\mathrm{BF}}
=\hat{\mathbf y}^{\mathsf T}
\bm\alpha^{\mathrm{cart}}\hat{\mathbf z}$.
This frame lets us compare the same molecular components along
the torsional path and examine their symmetry under well exchange
(Sec.~\ref{subsec:selection}). Below, we omit the superscript
$\mathrm{BF}$ on molecular components.
PBE0 calculations at the MP2 geometries give the dipoles and static polarizabilities, with the same basis sets and pseudopotential and a level-4 integration grid.\cite{Adamo1999}
The static polarizability is the field Hessian $\alpha_{ij}=-\partial^2 E/\partial F_i\partial F_j|_{\mathbf F=0}$, which we evaluate by central differences with field step $f=2\times10^{-4}$~a.u.
For an off-diagonal component, we combine energies computed with fields $\pm f$ along both axes $i$ and $j$:
\begin{align}
\alpha_{ij}\simeq
-\frac{E_{++}-E_{+-}-E_{-+}+E_{--}}{4f^2},
\label{eq:mixedstencil}
\end{align}
where $E_{st}=E(sf\hat e_i+tf\hat e_j)$ and $s,t=\pm1$.
For a diagonal component we use $-[E(f\hat e_i)-2E(0)+E(-f\hat e_i)]/f^2$.
The truncation error scales as $f^2$; at selected geometries, steps of $10^{-4}$ and $2\times10^{-4}$~a.u.\ agree to five significant figures.
We compute and transform the tensors at $+\phi$ and $-\phi$ separately, check their parity (Sec.~\ref{subsec:ramanresults}), and then impose parity when interpolating them for the nuclear matrix elements.

\subsection{Dipole and polarizability selection rules}
\label{subsec:selection}

We now determine which molecular properties can connect the
lowest even and odd torsional states. Two symmetries are
important: the $C_2$ rotation, which restricts the allowed
molecular components, and reflection, which determines whether
those components can couple states of opposite parity.

At each torsional angle $\phi$, the electric dipole
$\boldsymbol{\mu}(\phi)$ is a vector, whereas the electronic
polarizability $\boldsymbol{\alpha}(\phi)$ is a $3\times3$
tensor. Each tensor component, such as $\alpha_{yz}(\phi)$,
is a scalar value at that geometry.

A $180^\circ$ rotation about the molecular $x$ axis leaves
the geometry unchanged but reverses the $y$ and $z$ directions:
\begin{equation}
(x,y,z)\xrightarrow{C_2}(x,-y,-z).
\label{eq:C2axes}
\end{equation}
Because the molecule is unchanged, its dipole and polarizability
must also remain unchanged. The dipole components along $y$
and $z$ reverse sign under this rotation and therefore must
vanish:
\begin{equation}
\mu_y(\phi)=\mu_z(\phi)=0.
\label{eq:C2dipole}
\end{equation}
The polarizability components $\alpha_{xy}$ and $\alpha_{xz}$
also reverse sign and must vanish. However, $\alpha_{yz}$
is allowed because both $y$ and $z$ reverse, leaving their
product unchanged. Thus, the polarizability tensor has the form
\begin{equation}
\boldsymbol{\alpha}(\phi)=
\begin{pmatrix}
\alpha_{xx}(\phi)&0&0\\
0&\alpha_{yy}(\phi)&\alpha_{yz}(\phi)\\
0&\alpha_{yz}(\phi)&\alpha_{zz}(\phi)
\end{pmatrix}.
\label{eq:C2alpha}
\end{equation}

We next consider reflection in the molecular $xz$ plane,
which exchanges the two mirror-related torsional wells:
\begin{equation}
(x,y,z;\phi)\xrightarrow{\sigma_{xz}}
(x,-y,z;-\phi).
\label{eq:reflectionaxes}
\end{equation}
Under reflection, the $x$ and $z$ directions remain unchanged,
while $y$ reverses. Consequently, $\mu_x$ and the diagonal
polarizabilities have the same values in the two wells,
whereas $\alpha_{yz}$ changes sign:
\begin{align}
\mu_x(-\phi)&=\mu_x(\phi),
\label{eq:mueven}\\
\alpha_{ii}(-\phi)&=\alpha_{ii}(\phi),
\quad i=x,y,z,\\
\alpha_{yz}(-\phi)&=-\alpha_{yz}(\phi).
\label{eq:alphaodd}
\end{align}
Thus, $\mu_x$ and the diagonal polarizabilities are even
functions of $\phi$, while $\alpha_{yz}$ is odd.

These symmetry properties determine the torsional transition
matrix elements. A component such as $\alpha_{yz}(\phi)$ is one
number at a fixed geometry. For a torsional wavefunction spanning
many angles, its angle dependence defines a multiplication
operator:
\begin{equation}
(\hat O\psi)(\chi)=O[\phi(\chi)]\psi(\chi),
\end{equation}
where $O$ may be $\mu_x$ or one component of
$\boldsymbol{\alpha}$. The two lowest torsional eigenstates
have opposite parity:
$\psi_+(-\chi)=\psi_+(\chi)$ and
$\psi_-(-\chi)=-\psi_-(\chi)$.

For an even molecular property, the product
$\psi_+^*(\chi)O[\phi(\chi)]\psi_-(\chi)$ is odd.
Its integral over the symmetric torsional interval therefore
vanishes. This gives
\begin{align}
\mu_{01}^{\mathrm{tor}}
&=\langle+|\hat\mu_x|-\rangle=0,
\label{eq:dipoleforbidden}\\
\langle+|\hat\alpha_{ii}|-\rangle&=0,
\quad i=x,y,z.
\end{align}
The diagonal polarizabilities are not themselves zero;
only their transition matrix elements between the even
and odd states vanish.

The situation is different for $\alpha_{yz}$. Since both
$\alpha_{yz}(\phi)$ and $\psi_-(\chi)$ are odd,
their product is even. Contributions from the two
mirror-related wells therefore add rather than cancel.
The corresponding transition matrix element is
\begin{align}
A_{01}^{yz}
&=\langle+|\hat\alpha_{yz}|-\rangle\notag\\
&=\int_{-\pi}^{\pi}
\psi_+^*(\chi)\alpha_{yz}[\phi(\chi)]
\psi_-(\chi)\,d\chi.
\label{eq:ramanallowed}
\end{align}
Reflection symmetry permits a nonzero value of this molecular-frame
torsional integral. We retain the phase convention defined in
Sec.~\ref{subsec:dvr} and report its magnitude when comparing molecules.

This matrix element also has a direct interpretation in
terms of the localized torsional states. Since the expectation
value of an odd property vanishes in either parity eigenstate,
the localized states defined in Eq.~\eqref{eq:LR} satisfy
\begin{equation}
A_{01}^{yz}
=-\langle\mathrm L|\hat\alpha_{yz}|\mathrm L\rangle
= \langle\mathrm R|\hat\alpha_{yz}|\mathrm R\rangle.
\label{eq:identityLR}
\end{equation}
Thus, the two localized handed states have equal and opposite
average values of $\alpha_{yz}$. These averages include the
torsional wavefunction over each well and need not equal
the polarizability at the well minimum.

Therefore, although the lowest even--odd transition is
forbidden through the molecular-frame electric dipole,
it is symmetry-allowed through $\alpha_{yz}$. The calculated
$A_{01}^{yz}$ measures this internal torsional coupling;
the observability of a Raman line additionally depends on
the full rotation--torsion selection rules.

\subsection{Contributions along the torsional path}
\label{subsec:contributions}

The electric dipole $\mu_x(\phi)$ and polarizability
$\alpha_{yz}(\phi)$ change with the torsional angle as
the molecular geometry changes. Each angle therefore
contributes a different value of the molecular property
$O(\phi)$, where $O=\mu_x$ or $\alpha_{yz}$, to the
transition matrix element. The contribution also depends
on the even and odd torsional wavefunctions at that angle.

To identify which angles contribute most strongly,
we plot the signed contribution per degree:
\begin{equation}
\mathcal I_j^O=\psi_+(\chi_j)\,O(\phi_j)\,\psi_-(\chi_j)\,
\frac{\Delta\chi}{\Delta\phi_j^{(\mathrm{deg})}},
\label{eq:contributiondensity}
\end{equation}
where $\Delta\chi$ is the uniform DVR grid spacing and
$\Delta\phi_j^{(\mathrm{deg})}$ is the angular width
represented by grid point $j$, in degrees.

\subsection{Controlled isotope calculation}\label{subsec:isotope}

Both isotopologues use the same relaxed H$_2$O$_2$ geometries, CCSD(T)//MP2 potential, and PBE0 polarizability:
\begin{align}
V_{\mathrm{D_2O_2}}(\phi)&=V_{\mathrm{H_2O_2}}(\phi),\\
\alpha_{yz}^{\mathrm{D_2O_2}}(\phi)
&=\alpha_{yz}^{\mathrm{H_2O_2}}(\phi).
\end{align}
The two calculations differ only in the hydrogen mass; we use $m_{\mathrm H}=1.00782503223$, $m_{\mathrm D}=2.01410177812$, and $m_{\mathrm O}=15.99491461957$~amu.
Replacing H with D changes the center of mass, the rotational projection, $I_{\mathrm{eff}}(\phi)$, and $\chi(\phi)$.
We recompute Eqs.~\eqref{eq:omega}--\eqref{eq:arc} for both isotopologues with identical settings and no fit to D$_2$O$_2$ spectra.

OpenAI ChatGPT (GPT-6 Sol) assisted with literature review, theoretical interpretation, checks of derivations and numerical comparisons, and manuscript editing; the authors evaluated its suggestions against the calculations and sources.

\section{Results and discussion}\label{sec:results}

\subsection{Chemical trends in tunneling and coupling}\label{subsec:landscapes}

\begin{figure*}[!t]
\centering
\includegraphics[width=0.98\textwidth]
{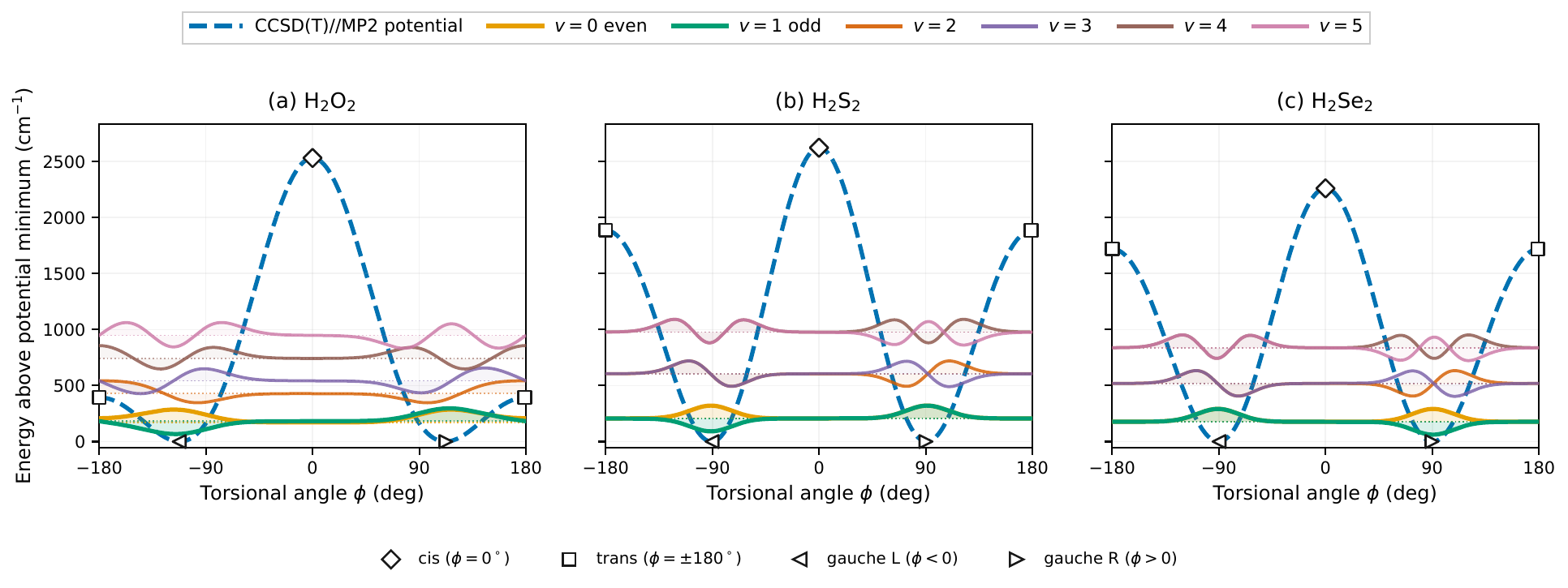}
\caption{\label{fig:landscapes}
Torsional potentials and six lowest eigenstates of (a) H$_2$O$_2$, (b) H$_2$S$_2$, and (c) H$_2$Se$_2$. Blue dashed curves show the relaxed CCSD(T)//MP2 potentials. All panels use the same linear energy scale, with zero at each potential minimum. Colored curves show signed wavefunctions centered on their calculated eigenenergies and rescaled for visibility. Ground-doublet splittings are listed in Table~\ref{tab:tunneling}.}
\end{figure*}

Table~\ref{tab:tunneling} and Fig.~\ref{fig:landscapes} summarize the potentials and kinetic constants across the series.
The cis/trans barriers are $2531.8/394.7~\mathrm{cm}^{-1}$ for H$_2$O$_2$, $2623.3/1887.0~\mathrm{cm}^{-1}$ for H$_2$S$_2$, and $2260.2/1720.5~\mathrm{cm}^{-1}$ for H$_2$Se$_2$.
The constant arc-coordinate inertia $I_{\mathrm{arc}}$ rises from
$1.503$ to $3.155$ and $3.860~\mathrm{amu\,bohr^2}$ across
H$_2$O$_2$, H$_2$S$_2$, and H$_2$Se$_2$.

The ground splitting falls from $10.984~\mathrm{cm}^{-1}$ in H$_2$O$_2$ to $5.636\times10^{-6}~\mathrm{cm}^{-1}$ in H$_2$S$_2$ and $1.743\times10^{-6}~\mathrm{cm}^{-1}$ in H$_2$Se$_2$, a $6.3\times10^6$-fold decrease.
The tunneling time grows accordingly, from $1.518~\mathrm{ps}$ to $9.57~\mu\mathrm{s}$.
Yet over the same series, $|A_{01}^{yz}|$ increases by $59\%$, from $1.57$ to $2.49$~a.u.\ (Table~\ref{tab:raman}).
Sections~\ref{subsec:excitedpairs} and~\ref{subsec:isotoperesult} test whether this contrast holds when the electronic surfaces are kept fixed.

\begin{table*}[!t]
\centering
\caption{\label{tab:tunneling}
Relaxed torsional barriers and ground-doublet properties. The cis and trans barriers are measured above each potential minimum. $I_{\mathrm{arc}}$ is the constant inertia defined in Eq.~\eqref{eq:arc}, and $B_{\mathrm{arc}}=\hbar^2/(2I_{\mathrm{arc}}hc)$ is its kinetic coefficient. $\Delta_{01}$ is the doublet splitting, $\nu_{01}=c\Delta_{01}$ is the corresponding frequency, and $t_{\mathrm L\rightarrow\mathrm R}=1/(2\nu_{01})$ is the first coherent well-to-well transfer time. D$_2$O$_2$ uses the same potential as H$_2$O$_2$ with deuterium masses.}
\begin{ruledtabular}
\begin{tabular}{lccccccc}
Molecule & $V_{\mathrm{cis}}$ & $V_{\mathrm{trans}}$
& $I_{\mathrm{arc}}$ & $B_{\mathrm{arc}}$ & $\Delta_{01}$ & $\nu_{01}$
& $t_{\mathrm L\rightarrow\mathrm R}$\\
& (cm$^{-1}$) & (cm$^{-1}$) & (amu bohr$^2$)
& (cm$^{-1}$) & (cm$^{-1}$) & & \\
\hline
H$_2$O$_2$ & 2531.8 & 394.7 & 1.503 & 40.1
& 10.984 & 329.3 GHz & 1.518 ps\\
D$_2$O$_2$ & 2531.8 & 394.7 & 2.784 & 21.6
& 1.628 & 48.81 GHz & 10.24 ps\\
H$_2$S$_2$ & 2623.3 & 1887.0 & 3.155 & 19.1
& $5.636\times10^{-6}$ & 169.0 kHz & $2.959~\mu$s\\
H$_2$Se$_2$ & 2260.2 & 1720.5 & 3.860 & 15.6
& $1.743\times10^{-6}$ & 52.3 kHz & $9.57~\mu$s\\
\end{tabular}
\end{ruledtabular}
\end{table*}

\subsection{Dipole and polarizability coupling}\label{subsec:ramanresults}

\begin{figure*}[!t]
\centering
\includegraphics[width=0.82\textwidth]
{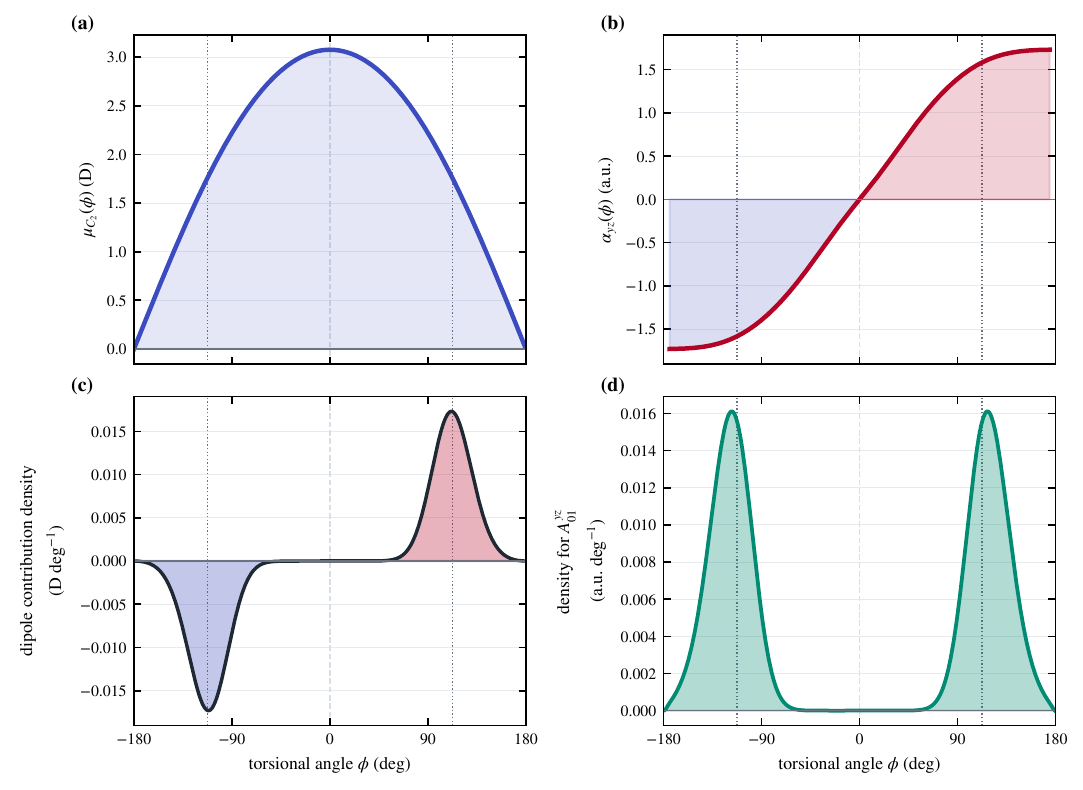}
\caption{\label{fig:operatorcontrast} Dipole and polarizability contributions for the H$_2$O$_2$ ground doublet. (a) The $C_2$-axis dipole is largest at cis. (b) $\alpha_{yz}$ changes sign between the wells. (c) Dipole contributions from the two wells cancel; the density vanishes at cis because the odd torsional wavefunction is zero there. (d) Polarizability contributions from the two wells add to give $A_{01}^{yz}$.}
\end{figure*}

The computed permanent dipole points along the $C_2$ axis and is even under well exchange, so its pure torsional matrix element between opposite-parity states vanishes.
The odd $\alpha_{yz}$ component does couple these states (Fig.~\ref{fig:operatorcontrast}).
This torsional selection rule does not forbid infrared rotation--tunneling transitions, which also involve rotational selection rules and space-fixed dipole components.\cite{Hougen1984,Flaud1989,AlRefaie2016}

Before imposing symmetry, we compared the independently calculated
polarizabilities at $+\phi$ and $-\phi$. The largest value of
$|\alpha_{yz}(-\phi)+\alpha_{yz}(\phi)|$ over geometries where the
frame is defined was $1.6\times10^{-5}$, $6.1\times10^{-5}$, and
$3.3\times10^{-5}$~a.u.\ for H$_2$O$_2$, H$_2$S$_2$, and H$_2$Se$_2$,
respectively.

\begin{figure*}[!t]
\centering
\includegraphics[width=0.85\textwidth]
{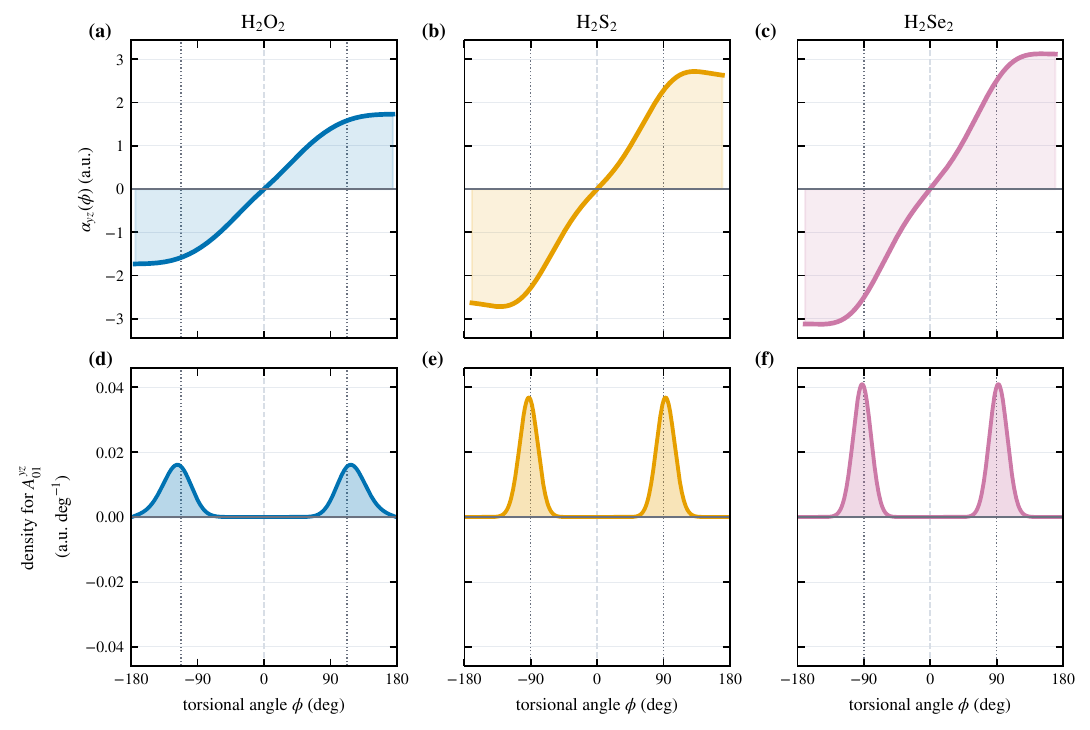}
\caption{\label{fig:ramanmechanism}
Where the $yz$ polarizability matrix element forms in H$_2$O$_2$,
H$_2$S$_2$, and H$_2$Se$_2$. (a)--(c) Odd $\alpha_{yz}(\phi)$ profiles. (d)--(f) Contribution
densities [Eq.~\eqref{eq:contributiondensity}]. Each integrates to $A_{01}^{yz}$ in
Table~\ref{tab:raman}.}
\end{figure*}

Table~\ref{tab:raman} lists $|\alpha_{yz}|$ at each torsional minimum ($\phi=112.2^\circ$, $90.7^\circ$, and $90.2^\circ$ for O, S, and Se).
It grows from $1.58$~a.u.\ in H$_2$O$_2$ to $2.30$ in H$_2$S$_2$ and $2.52$ in H$_2$Se$_2$, and $|A_{01}^{yz}|$ follows it: in each molecule of the chemical series, $|A_{01}^{yz}|$ is about $1\%$ smaller than the equilibrium value.
Equation~\eqref{eq:identityLR} identifies $A_{01}^{yz}$ as a localized-state average. Figure~\ref{fig:ramanmechanism} shows that most of its contribution comes from the wells, whereas the splitting depends on barrier penetration.

\begin{table*}[!t]
\centering
\caption{\label{tab:raman}
Ground-doublet splitting and polarizability matrix element compared with
$\alpha_{yz}$ at the torsional minimum. H$_2$O$_2$ and D$_2$O$_2$ share the
electronic surface and equilibrium angle.}
\begin{ruledtabular}
\begin{tabular}{lcccc}
Molecule & $\Delta_{01}$ (cm$^{-1}$)
& $|\alpha_{yz}(\phi_{\mathrm{eq}})|$ (a.u.)
& $|A_{01}^{yz}|$ (a.u.)
& $|A_{01}^{yz}|/|\alpha_{yz}(\phi_{\mathrm{eq}})|$\\
\hline
H$_2$O$_2$  & 10.984 & 1.5814 & 1.5652 & 0.9898\\
D$_2$O$_2$  & 1.628 & 1.5814 & 1.5811 & 0.9998\\
H$_2$S$_2$  & $5.636\times10^{-6}$ & 2.2985 & 2.2715 & 0.9883\\
H$_2$Se$_2$ & $1.743\times10^{-6}$ & 2.5154 & 2.4941 & 0.9915\\
\end{tabular}
\end{ruledtabular}
\end{table*}

\subsection{Excited torsional doublets}\label{subsec:excitedpairs}

Within a molecule, the three lowest even--odd pairs ($n=0,1,2$; Fig.~\ref{fig:landscapes}) share the potential, polarizability surface, and masses and differ only in torsional excitation. Table~\ref{tab:excitedpairs} lists their splittings $\Delta_n$ and squared couplings $S_n^{yz}$.

\begin{table*}[!t]
\centering
\caption{\label{tab:excitedpairs}
Three lowest parity pairs: splitting $\Delta_n$ and squared coupling
$S_n^{yz}=|\langle n,+|\hat\alpha_{yz}|n,-\rangle|^2$. The last column is the
ratio to the ground pair of the same molecule.}
\begin{ruledtabular}
\begin{tabular}{lcccc}
Molecule & $n$ & $\Delta_n$ (cm$^{-1}$)
& $S_n^{yz}$ (a.u.$^2$) & $S_n^{yz}/S_0^{yz}$\\
\hline
H$_2$O$_2$ & 0 & 10.984 & 2.450 & 1.000\\
& 1 & 114.346 & 1.787 & 0.730\\
& 2 & 206.250 & 1.070 & 0.437\\
D$_2$O$_2$ & 0 & 1.628 & 2.500 & 1.000\\
& 1 & 38.365 & 2.163 & 0.865\\
& 2 & 120.845 & 1.388 & 0.555\\
H$_2$S$_2$ & 0 & $5.636\times10^{-6}$ & 5.160 & 1.000\\
& 1 & $5.64\times10^{-4}$ & 4.914 & 0.952\\
& 2 & 0.024689 & 4.693 & 0.910\\
H$_2$Se$_2$ & 0 & $1.743\times10^{-6}$ & 6.220 & 1.000\\
& 1 & $1.80\times10^{-4}$ & 6.000 & 0.965\\
& 2 & 0.008069 & 5.788 & 0.930\\
\end{tabular}
\end{ruledtabular}
\end{table*}

H$_2$Se$_2$ gives the clearest case.
For coherently prepared localized states, the tunneling time falls from $9.57~\mu\mathrm{s}$ ($n=0$) to $92.7~\mathrm{ns}$ ($n=1$) and $2.07~\mathrm{ns}$ ($n=2$), while $S_n^{yz}$ falls by only $3.5\%$ and $7.0\%$.
H$_2$S$_2$ shows the same pattern: its excited pairs have splittings about $100$ and $4.4\times10^3$ times larger than the ground pair, yet keep $95.2\%$ and $91.0\%$ of the ground-pair squared coupling.
In both molecules, all three pairs have $P_{\mathrm{well},n}>0.99999$, so the small loss of coupling reflects how each state samples $\alpha_{yz}$ within one well.

The oxygen pairs are less localized.
The excited pairs of H$_2$O$_2$ and D$_2$O$_2$ have $P_{\mathrm{well},n}$ between $0.863$ and $0.962$, compared with $0.9938$ and $0.9991$ for their ground pairs.
Part of each excited wavefunction reaches the opposite well, where $\alpha_{yz}$ has the opposite sign and cancels some of the contribution from the original well; this partly explains their weaker coupling.
The well-to-well picture of the excited pairs therefore holds best for S and Se.

\subsection{Deuteration isolates the mass effect}\label{subsec:isotoperesult}

Replacing H with D on the same potential and polarizability surfaces isolates the effect of nuclear mass.
The constant arc-coordinate inertia $I_{\mathrm{arc}}$ rises from
$1.5025$ to $2.7841~\mathrm{amu\,bohr^2}$ upon deuteration.
The ground splitting falls from $10.984$ to $1.628~\mathrm{cm}^{-1}$, a factor of $6.75$, and the tunneling time rises from $1.518$ to $10.24~\mathrm{ps}$.
Yet $|A_{01}^{yz}|$ changes only from $1.5652$ to $1.5811$~a.u., or $1.01\%$.
Both isotopologues share $|\alpha_{yz}(\phi_{\mathrm{eq}})|=1.5814$~a.u., and the ratio $|A_{01}^{yz}|/|\alpha_{yz}(\phi_{\mathrm{eq}})|$ rises from $0.9898$ to $0.9998$, consistent with the better-localized D$_2$O$_2$ ground pair ($P_{\mathrm{well},0}=0.9991$ versus $0.9938$).
Deuteration thus slows tunneling $6.75$-fold but barely changes the in-well average of $\alpha_{yz}$.

\subsection{Experimental and theoretical benchmarks}\label{subsec:validation}

Table~\ref{tab:validation} compares the calculated ground splittings with experimental assignments and independent calculations, and Table~\ref{tab:origins} lists the H$_2$O$_2$ and D$_2$O$_2$ torsional origins.

\begin{table}[!t]
\centering
\caption{\label{tab:validation}
Ground-doublet splittings $\Delta_{01}$ (cm$^{-1}$) compared with independent
values. Oxygen entries are experimental
assignments;\cite{Flaud1989,HuntLeacock1966} S and Se entries are instanton estimates.\cite{Sahu2021}}
\begin{ruledtabular}
\begin{tabular}{lccl}
Molecule & DVR & Literature & Type\\
\hline
H$_2$O$_2$  & 10.984 & 11.4372 & expt.\\
D$_2$O$_2$  & 1.628 & 1.88 & expt.\\
H$_2$S$_2$  & $5.636\times10^{-6}$ & $4.3\times10^{-6}$ & instanton\\
H$_2$Se$_2$ & $1.743\times10^{-6}$ & $1.5\times10^{-6}$ & instanton\\
\end{tabular}
\end{ruledtabular}
\end{table}

\begin{table}[!t]
\centering
\caption{\label{tab:origins}
Torsional levels (cm$^{-1}$) of H$_2$O$_2$ and D$_2$O$_2$ relative to the ground state (level 0). Literature values are from Flaud et al.\cite{Flaud1989} for H$_2$O$_2$ and from Hunt and Leacock\cite{HuntLeacock1966} for D$_2$O$_2$.}
\begin{ruledtabular}
\begin{tabular}{lrrr}
Molecule & Level & Calculated & Literature values\\
\hline
H$_2$O$_2$ & 1 & 10.984 & 11.4372\\
 & 2 & 257.3 & 254.5499\\
 & 3 & 371.7 & 370.8932\\
 & 4 & 571.0 & 569.7442\\
 & 5 & 777.3 & 776.1215\\
D$_2$O$_2$ & 1 & 1.628 & 1.88\\
 & 2 & 215.0 & 208.6\\
 & 3 & 253.4 & 250.9\\
 & 4 & 392.7 & 387.7\\
 & 5 & 513.6 & 511.2\\
\end{tabular}
\end{ruledtabular}
\end{table}

For H$_2$O$_2$, the rms deviation over the five excited levels is $1.51~\mathrm{cm}^{-1}$.
The ground splitting is $4.0\%$ below the assignment of Flaud et al., and the first excited pair splits by $114.35~\mathrm{cm}^{-1}$ versus the assigned $116.34~\mathrm{cm}^{-1}$.\cite{Flaud1989}
The fitted potential of Flaud et al.\ has cis/trans barriers of $2562.8/387.07~\mathrm{cm}^{-1}$ and a torsional angle of $111.9\pm0.4^\circ$; our barriers differ by $-1.2\%/+2.0\%$, and our angle of $112.2^\circ$ lies within this range.
The six-dimensional calculation of Fehrensen et al.\ on the PCPSDE surface gives a ground splitting of $11.0~\mathrm{cm}^{-1}$ and next origins at $259.3$ and $378.2~\mathrm{cm}^{-1}$;\cite{Fehrensen2007} our values of $10.984$, $257.3$, and $371.7~\mathrm{cm}^{-1}$ agree to within about $2\%$.
Because the electronic surfaces differ, we cannot attribute the remaining differences to dimensionality alone.

For D$_2$O$_2$, the rms deviation over the five nonzero origins
inferred by Hunt and Leacock is $3.95~\mathrm{cm}^{-1}$.
\cite{HuntLeacock1966}
The first excited pair splits by $38.37~\mathrm{cm}^{-1}$ versus $42.3~\mathrm{cm}^{-1}$.
We underestimate the ground splitting by $13.4\%$, but the four higher origins lie within $3.1\%$ of their assignments.

Quadruple-zeta single-point calculations at cis, trans, and one nonplanar reference geometry shift the sampled relative energies by up to $18.3$, $82.8$, and $119.1~\mathrm{cm}^{-1}$ for O, S, and Se. These checks show basis sensitivity at the sampled points but cannot determine the correction to the full potential or its tunneling splitting.

For H$_2$S$_2$ and H$_2$Se$_2$, the CCSD(T)-corrected trans-path instanton estimates of Sahu, Richardson, and Berger are $4.3\times10^{-6}$ and $1.5\times10^{-6}~\mathrm{cm}^{-1}$;\cite{Sahu2021} our values are $31\%$ and $16\%$ higher.
Because the paths and tunneling methods differ, this comparison does not measure our accuracy.
H$_2$S$_2$ rotational spectra show torsion-related line doublings near $150~\mathrm{kHz}$,\cite{Mittler1990,WinnewisserYamada1991} but their dependence on rotation prevents a direct comparison with our rotation-free ground-doublet frequency of $169~\mathrm{kHz}$.

\subsection{Implications and limitations}\label{subsec:scope}

The nonzero $A_{01}^{yz}$ is an intrinsic torsional coupling that can enter a Raman transition.
Observed Raman intensities and polarizations also depend on molecular rotation, state populations, nuclear-spin weights, and the light geometry.\cite{Placzek1934,Long2002}
Pitsevich et al.\ calculated rotation--torsion Raman lines for H$_2$O$_2$ and found weak pure-torsional features in their model.\cite{Pitsevich2010}
Ordinary Raman intensities are the same for both enantiomers; distinguishing them requires a chiroptical measurement.\cite{BarronBuckingham1971}

The model has three main limitations.
The reduced path omits coupling to other vibrations, their
torsion-dependent zero-point energies, and explicit rotation--torsion
dynamics. The static PBE0 polarizabilities have not been benchmarked
against correlated response calculations or dynamic polarizabilities
at optical frequencies. The transfer times describe isolated coherent
motion and do not include decoherence or thermal transitions.

\section{Conclusions}\label{sec:conclusions}

Within this torsional model, symmetry forbids an electric-dipole transition between the opposite-parity states of a doublet, while the odd polarizability component $\alpha_{yz}$ gives a nonzero transition matrix element. Across H$_2$O$_2$, H$_2$S$_2$, and H$_2$Se$_2$, the ground-doublet splitting decreases by a factor of $6.3\times10^6$, yet $|A_{01}^{yz}|$ increases from $1.57$ to $2.49$~a.u. The polarizability coupling therefore does not diminish as coherent tunneling slows.

Two comparisons on fixed electronic surfaces make this separation more compelling. In H$_2$Se$_2$, three well-localized doublets span first well-to-well transfer times from $2.07~\mathrm{ns}$ to $9.57~\mu\mathrm{s}$ while retaining at least $93\%$ of the ground-doublet squared coupling; H$_2$S$_2$ shows the same pattern. Deuteration slows ground-doublet tunneling in hydrogen peroxide by a factor of $6.75$ but changes its transition polarizability by only $1.01\%$. The contribution densities reveal the mechanism: the polarizability matrix element forms mainly inside the nonplanar wells, whereas the splitting is sensitive to penetration through the barriers between them. Coherent oscillations of torsional handedness can thus change by orders of magnitude without a corresponding loss of intrinsic polarizability coupling, providing a physical basis for exploring Raman-based control of chiral torsional motion.

\begin{acknowledgments}
S.K. would like to acknowledge the support of the U.S. Department of Energy (DOE) Office of Basic Energy Sciences, under grant number DESC0026309. 
M.S. would like to acknowledge the use of resources of the Oak Ridge Leadership Computing Facility at the Oak Ridge National Laboratory, which is supported by the Office of Science of the US Department of Energy under Contract No. DEAC05-00OR22725. This manuscript has in part been authored by UT-Battelle, LLC under Contract No. DE-AC05-00OR22725 with the U.S. Department of Energy. The United States Government retains and the publisher, by accepting the article for publication, acknowledges that the US Government retains a nonexclusive, paid-up, irrevocable, worldwide license to publish or reproduce the published form of the manuscript, or allow others to do so, for US Government purposes. The Department of Energy will provide public access to these results of federally sponsored research in accordance with the DOE Public Access Plan (http://energy.gov/downloads/doe-publicaccess-plan).
\end{acknowledgments}

\section*{Author declarations}

\subsection*{Conflict of interest}

The authors have no conflicts to disclose.

\subsection*{Author contributions}

\textbf{Shah Ishmam Mohtashim}: Conceptualization (equal); Methodology (lead);
Software (lead); Validation (lead); Formal analysis (lead); Investigation
(lead); Data curation (lead); Visualization (lead); Writing -- original draft
(lead); Writing -- review and editing (equal).
\textbf{Manas Sajjan}: Conceptualization (equal); Methodology (supporting);
Formal analysis (equal); Supervision (equal); Writing -- review and
editing (equal).
\textbf{Sabre Kais}: Conceptualization (equal); Formal analysis (supporting);
Supervision (lead); Project administration (lead); Funding acquisition (lead);
Writing -- review and editing (equal).

\section*{Data availability}

The data that support the findings of this study are openly available at \url{https://github.com/IshmamShah/DoubleWell}.

\bibliography{references}

\end{document}